\documentclass[amsmath, amssymb, breqn, aps, prl, superscriptaddress, twocolumn, longbibliography]{revtex4-2}

\usepackage{graphicx}
\usepackage{dcolumn}
\usepackage{bm}
\usepackage{hyperref}
\usepackage{comment}
\usepackage{float}

\def\4He{$^4$He}

\newcommand{\rhon}{\ensuremath{\rho_n}}

\newcommand{\bv}[1]{\ensuremath{\mathbf{#1}}}
\newcommand{\dd}{\ensuremath{\operatorname{d}}}

\begin{document}

\title{\textbf{Decay of two-dimensional superfluid turbulence over pinning surface}
}%

\author{Filip Novotný}
 \email{filip.novotny@mff.cuni.cz}
\author{Marek Talíř}%
\author{Emil Varga}
 \homepage{emil.varga@matfyz.cuni.cz}
\affiliation{Faculty of Mathematics and Physics, Charles University, Ke Karlovu 3, 121 16 Prague,
Czech Republic}%

\date{\today}

\begin{abstract}
We report on the free decay of quasi-two-dimensional turbulence in superfluid $^4$He confined within nanofluidic channels. Using a pump-probe technique, we observe a complex decay of the vortex density $L(t)$ that deviates from a simple power law. The decay exhibits a universal fast transient, scaling as $L\propto t^{-2}$, followed by a slower non-universal regime that depends on the geometry and flow conditions. We demonstrate that this behavior is governed by the interplay between vortex pinning on the disordered topography of the channel walls and the mobilizing effect of the weak probe flow. A numerical model that treats pinning as a velocity-dependent effective mutual friction successfully reproduces the essential features of our experimental observations.
\end{abstract}

\maketitle

\textit{Introduction -- }Two-dimensional turbulence is an idealization of real three-dimensional flow that is nevertheless a useful model of many real flows, such as large-scale motion in oceans and atmospheres \cite{hogstrom_1999,rosell-fieschi_2015}. Of particular interest is flow over disordered topography, which, depending on the available kinetic energy, can exist in the state of freely roaming vortices or vorticity locked to the topography in the case of geostrophic flows \cite{Bretherton1976,Siegelman2023,Gallet2024}. The study of decaying turbulence can help elucidate the dissipation mechanisms and turbulence spectra and has been one of the most fruitful lines of research in both classical and quantum turbulence \cite{QTbook,davidson_turbulence_2015}. In certain regimes, 2D turbulence resembles a vortex gas \cite{carnevale_1991,Gallet2020} for which quantized vortices in 2D superfluids are a paradigmatic example \cite{tilley_2003}. Previous studies of their decay in superfluid helium \cite{sachkou_2019} and Bose-Einstein condensate \cite{gauthier_2019} were limited by low vortex numbers and short observation times. Here, we study the decay of quasi-2D turbulence in superfluid helium within nanofluidic channels, starting from thousands of vortices observed over thousands of large eddy turnover times.

\begin{figure}
    \centering
    \includegraphics[width=\linewidth]{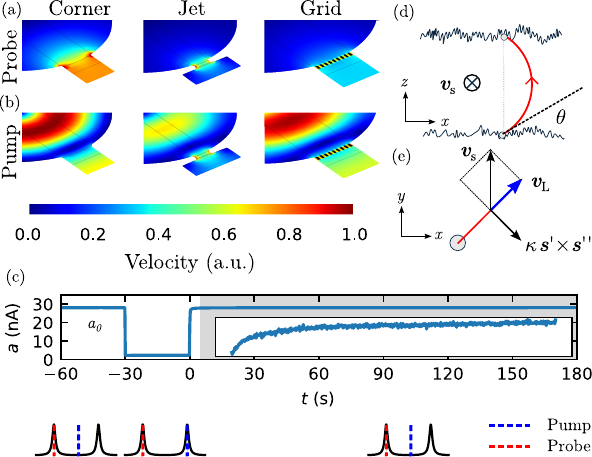}
    \caption{(a, b) Acoustic simulation of the superfluid resonant modes inside the nanofluidic cavity used in pump-probe measurement for the three used resonators. The frequencies are approximately 2~kHz for probe and 30~kHz for pump modes \cite{novotny_inter_2025}.  (c) A scheme of the measurement sequence, the bottom sketch shows the detuning of the probe and pump frequencies on the corresponding resonances. The inset axes show a zoomed view of $t > 5$~s data (shaded area in the main plot). (d,e) Vortex depinning model in the numerical simulation (see text).}
    \label{fig:meas_sequence}
\end{figure}

Superfluid turbulence can be characterized using the vortex line density $L$, i.e., the total length of vortex lines per unit volume. The vortex line density $L(t)$ typically decays as a power law $L(t)\propto t^{-\xi}$. While 3D superfluid turbulence exhibits $\xi=1$ (Vinen) or $\xi=3/2$ (Kolmogorov) \cite{QTbook,walmsley_2008,novotny_sphere_2024,babuin_2014,babuin_2015,zmeev_2015}, 2D Bose-Einstein condensates \cite{kwon_2014,stagg_2015,cidrim_2016,seo_2017,baggaley_decay_2018,kanai_2021} at zero temperature show $\xi=1/2$ and $\xi=1/3$, respectively, for the three- and four-vortex annihilation processes \cite{baggaley_decay_2018}). At finite temperatures, dissipation via mutual friction enables direct two-vortex annihilation predicting $\dd L/\dd t = -\Gamma_2 L^2(t)$, where $\Gamma_2$ is a vortex annihilation rate, i.e., for $L(t=0)\gg 1$, $L(t) = 1/(\Gamma_2 t)$ and $\xi=1$ \cite{baggaley_decay_2018}. In superfluid helium above 1~K the $L(t)\propto t^{-1}$ decay from two-vortex annihilation should dominate. To test this hypothesis, we use the nanofluidic Helmholtz resonators that were used in the past to study confined superfluidity \cite{rojas_2015,souris_2017,shook_stabilized_2020,Shook2024} and steady-state turbulence in oscillatory flows \cite{varga_2020,novotny_critical_2025}. The nanofluidic Helmholtz resonators used in this study are formed from a cavity of height $D\approx500$~nm enclosed by two fused silica chips (see refs.~\cite{rojas_2015,souris_2017} for fabrication details). The nanofluidic cavity consists of a circular ,,basin'' connected to the surrounding bath by two opposing channels (see Fig.~\ref{fig:meas_sequence}a and Fig.~\ref{fig:circuit}a in End Matter). The acoustic modes in these devices are driven and sensed via electrodes deposited on the top and bottom walls of the nanofluidic cavity, forming a parallel plate capacitor \cite{rojas_2015,souris_2017,varga_electromechanical_2021} (see the End Matter for details).

The acoustic motion can transition to turbulence at sufficiently high velocities \cite{varga_2020,novotny_critical_2025,novotny_inter_2025}. The transition to turbulence in these devices shows a remarkable complexity involving multiple large-scale turbulent states with stochastic transitions, which are related to the excitation of progressively larger modes via the inverse turbulent cascade \cite{varga_2020,novotny_critical_2025}. In previous experiments \cite{varga_2020,novotny_critical_2025}, the forcing and probing of the turbulence were combined into a single acoustic mode. Here, we use a two-mode pump-probe approach \cite{novotny_inter_2025}: a high-amplitude radial overtone (Fig.~\ref{fig:meas_sequence}b) generates turbulence predominantly in the inlet channels \cite{novotny_inter_2025}, while the fundamental Helmholtz mode (Fig.~\ref{fig:meas_sequence}a) continuously probes the vortex line density $L(t)$ via its attenuation. The probe mode is continuously driven below the critical velocity for onset of turbulence and its response is recorded as a function of time. The turbulence is switched on by tuning the pump frequency to its resonance and switched off by detuning it to maintain a constant RMS voltage on the basin capacitor. The typical measurement sequence (Fig.~\ref{fig:meas_sequence}c) consists of 30 s of quiescent state, followed by 30 s of turbulent forcing, and 180 s of decay, with 30 to 100 decay events ensemble averaged per parameter set. From the averaged fourth sound signal $a(t)$ the vortex density is calculated as \cite{novotny_rot_2024}
\begin{equation}
    \label{eq:L}
    L(t) = \dfrac{2 \pi \Delta_0}{\alpha \kappa} \left( \dfrac{a_0}{a(t)} - 1 \right),
\end{equation}
where $a_0$ is an unattenuated probe signal taken as the average from first 25~s of the initial quiescent state, $\Delta_0$ the unattenuated probe peak width, $\alpha$ the mutual friction coefficient \cite{donnelly_1998} and $\kappa = 9.97 \times10^{-8}$~m$^2$s$^{-1}$ is the quantum of circulation. This relation was validated using a rotating cryostat \cite{novotny_rot_2024}, although in the present case $L$ is likely renormalized by dissipation due to pinning (see discussion). For non-uniform distribution of vortex line density, \eqref{eq:L} gives the average weighted by the velocity distribution of the acoustic mode \cite{novotny_inter_2025}.

We use three different Helmholtz resonators with a nanofluidic cavity height of $D\approx500$~nm, identical to those used in the steady state study \cite{novotny_critical_2025}, which differ in the geometry of the junction of channel and basin. The vortices in this system can be considered point-like since the lowest Kelvin wave frequency is two orders of magnitude higher than the acoustic frequency \cite{yang_quantum_2025}. Following ref.~\cite{novotny_critical_2025} we denote ``C'' (corner) the resonator with wide channel, ``J'' (jet) a narrow channel and ``G''  a grid at the connection between the basin and the channels; see geometry of the resonator channels in Fig.\ref{fig:meas_sequence}a and Fig.~\ref{fig:circuit}b for exact dimensions.

\begin{figure}
    \centering
    \includegraphics{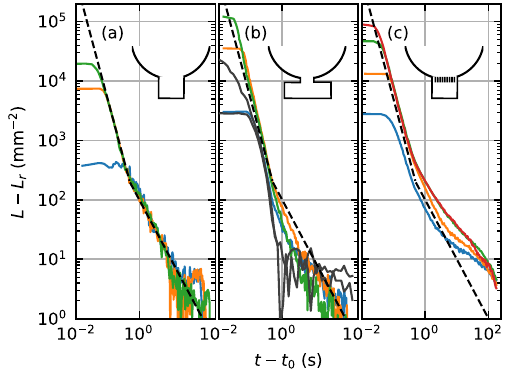}
    \caption{Vortex line density decay from all three geometries for a range initial densities $L_0$ and similar probe velocity amplitudes $v_p\approx 12$~cm/s for all, except for the two gray curves in panel (b), for which $v_p\approx 4.2$~cm/s. The dashed straight lines indicate $50/t^2$ (for $t < 0.5$~s) and $100/t$ (for $t > 0.5$~s) and are identical for all three. The delay on the order of 10~ms is consistent with the inverse linewidth of the 4th sound resonance.}
    \label{fig:decay-data-all}
\end{figure}

\emph{Results} - The decay starting from several initial vortex densities $L_0$ and probe velocity amplitudes $v_p$ for the three geometries is shown in Fig.~\ref{fig:decay-data-all} at 1.3~K (all experiments took place at saturated vapor pressure). The shown curves are filtered by adjacent averaging (see End Matter for the smoothing procedure). The velocity of the probe mode is calculated from the resonance amplitude, equivalent to \cite{novotny_critical_2025}, before the turbulence is generated. As $t=0$, the moment when the fourth sound attenuation begins to recover from the turbulent steady state is chosen. In Fig.~\ref{fig:decay-data-all}, the $x$ axis is shifted by the ,,virtual origin time'' $t_0$, chosen to maximize the extent of the power law scaling of the fast initial decay; $t_0 \approx 0 \pm 0.2$~s for all cases. Likewise, the $y$ axis is shifted by the remanent vorticity $L_r$, which arises since the calculation of $L$ using \eqref{eq:L} relies on the reference value $a_0$ which refers to a state with an unknown number of vortices. It was chosen again to maximize the extent of the late-time power law scaling or was kept zero if no power law scaling was present; $|L_r| < 5$~mm$^{-2}$ for all measured cases. Both $t_0$ and $L_r$ were chosen by hand.

All geometries show similar behavior: after the initial fast decay for $t \lesssim 0.5$~s a slow decay follows. The fast decay is described well with universal $L(t) = d_2/t^2$, with $d_2\approx 50$~mm$^{-2}$s$^2$ for sufficiently high $L_0$, which is suppressed or absent at lower initial densities $L_0$ (see Fig.~\ref{fig:decay-data-all}a,c bottom curves). For $t \gtrsim 0.5$~s ($L(t) \lesssim 4\times 10^2$~mm$^{-2}$) power law with an exponent close to $\xi = 1$ follows. For the C-type device, the $1/t$ decay independent of initial density $L_0$ is present for two orders of magnitude in both time and $L$, however, both J- and G-type devices show significant non-universal ($L_0$ dependent) deviations from self-similar decay. A limited dataset at 1.45~K shows similar behavior (see Fig.~\ref{fig:decays_G_1450}, End Matter). The experiments are limited to the lower end of the available temperatures due to the quality factor of the pump resonance rapidly decreasing with increasing temperature \cite{novotny_inter_2025}.

The exponent $\xi = 1$ corresponds to the vortex-antivortex pair annihilation, which was observed in experiments \cite{kwon_2014} and numerical simulations \cite{baggaley_decay_2018} in Bose-Einstein condensates with friction. This is surprising since the present flow is expected to be strongly influenced by vortex pinning. The pinning occurs on the roughness of the substrate (aluminum electrode), which is approximately 4~nm RMS (measured using atomic force microscopy \cite{novotny_critical_2025}). For vortices stretched between the top and bottom walls distance $D$ apart, the \emph{depinning velocity} $v_d$ needed to dislodge a vortex from a hemispherical bump of radius $b$ is $v_d =\kappa/(2 \pi D) \ln \left(b/a_0\right)$ \cite{schwarz_1985}. For $D = 500$~nm the height of the cavity, substituting the surface roughness for $b$, and $a_0 = 0.15$~nm the vortex core radius we get $v_d \approx 11$~cm/s. Note that the typical velocity induced by the vortex system $v_i \approx \kappa/(2\pi)\sqrt{L} \approx 0.5$~cm/s and thus vortices cannot depin under the superflow induced by nearest neighbors. The vortices must therefore become mobile due to the probe flow, which, for data in Fig.~\ref{fig:decay-data-all} had amplitude approximately 12~cm/s in the 1~mm section of the channel for devices C and G and in the narrowest section for device J.

\begin{figure}
    \centering
    \includegraphics{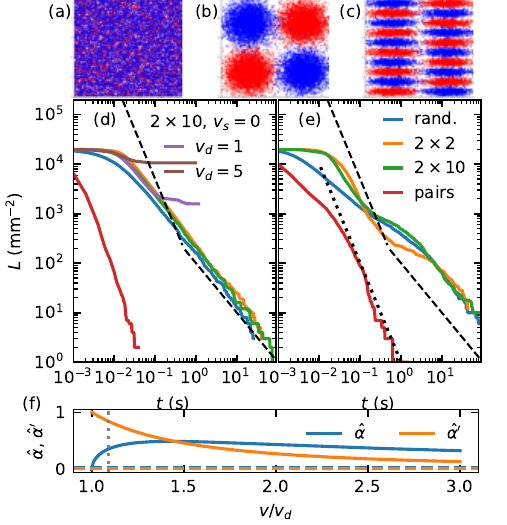}
    \caption{Decay of vortex density in the numerical simulation. (a-c) The initial conditions used in the simulations, from left to right: random or vortex pairs (see text), $2\times 2$ and $2 \times 10$. (d) The decay of the vortex density without probe flow: without pinning (lower four curves, same legend as in (e)) and with pinning (depinning velocity $v_d$ in the legend in cm/s). (e) The effect of initial conditions on the decay with $v_\mathrm{pin} = 10$~cm/s and $v_\mathrm{p} = 11$~cm/s. The dotted line shows $1[\mathrm{mm}^{-2}\mathrm{s}^2]/t^2$.
    (f) Effective mutual friction parameters. The dashed lines are bare $\alpha$ and $\alpha^\prime$ at 1.3~K \cite{donnelly_1998}. The gray dashed line corresponds to $v/v_\mathrm{d} = 12/11$.}
    \label{fig:simulations}
\end{figure}

\emph{Numerical model. -- }To account for the pinning and probe flow we need an effective equation of motion of a vortex moving between homogeneously rough surfaces. Assuming the critical angle model of depinning of vortices on rough surfaces originally considered by Lipniacki \cite{Lipniacki1997} and recently used in vortex filament simulations \cite{doyle_modelling_2024,nakagawa_dynamics_2023}, the vortex will start sliding along the wall if the contact angle $\theta < \theta_c$ for some critical angle $\theta_c$ (see Fig.~\ref{fig:meas_sequence}d).  For a pinned vortex ($\theta > \theta_c$), the contact angle is given by the balance between the self-induced velocity of a curved vortex and the externally imposed superflow of magnitude $v_s$. Assuming that the pinned vortex forms a section of a circle of radius $R$, the self-induced velocity in local-induction approximation is given by $v_i(R) = \kappa/(2\pi R)\ln\left(8R/a\right)$ and the vortex bows in the direction perpendicular to the applied superflow. The critical angle is equivalent to critical radius of curvature $R_c = D/(2\cos\theta_c)$. First, we consider the case without mutual friction of a depinned vortex subject to $v_s > v_d = v_i(R_c)$. The details of this sliding will depend in a complicated way on the microscopic topography of the rough wall. We further assume that (i) the vortex is always in its final equilibrium state for a given $v_s$ (i.e., Kelvin waves generated by repositioning of the pinning point are quickly attenuated and that the pinning points on top and bottom walls remain aligned); and (ii) the repositioning of the pinning point is dissipative, which momentarily shortens the vortex and thus increases the curvature radius $R$. This means that, momentarily, the applied superflow and the self-induced velocity are unbalanced, which will cause the vortex to rotate toward the direction of superflow, before it regrows and repositions. This rotation will cease once the azimuthal component of the applied velocity in the plane parallel with the flow is compensated by self-induced velocity of the vortex of magnitude $v_d$, see Fig.~\ref{fig:meas_sequence}e. After some trigonometry (see End Matter), this leads to vortex velocity without mutual friction
\begin{equation}
    \label{eq:vl}
    \bv{v}_L = (1 - \beta^2)\bv{v}_s - \beta\sqrt{1 - \beta^2}\bv{s}'\times\bv{v}_s,
\end{equation}
where $\beta = v_\mathrm{d}/|\bv{v}_s|$ and $\bv{s} = \pm \hat{\bv{z}}$ is the unit vector along direction of vorticity. This can be recast in terms of balance of forces on a massless vortex $\bv{f}_M + \bv{f}_D = 0$, where the Magnus force $\bv{f}_M = \rho_s\kappa\bv{s}'\times(\bv{v}_L - \bv{v}_s)$ and the pinning drag force $\bv{f}_D = -\rho_s\kappa \Gamma \bv{v}_L$ where $\Gamma = \beta/\sqrt{1 - \beta^2}$. Including the mutual friction $f_\mathrm{MF} = \gamma_0 (\bv{v}_n - \bv{v}_L) + \gamma_0'\bv{s}'\times(\bv{v}_n - \bv{v}_L)$, \cite{Barenghi1983}, and assuming $\bv{v}_n = 0$, the total force acting on a vortex is
\begin{equation}
    \label{eq:F-total}
    \bv{F}_\mathrm{v} = \rho_s\kappa \mathbf{s'}\times (\bv{v}_L - \bv{v}_s) -\rho_s\kappa\Gamma \bv{v}_L - \gamma_0\bv{v}_L - \gamma_0'\mathbf{s'}\times\bv{v}_L,
\end{equation}
and the final vortex velocity $\bv{v}_L$ is found from $\bv{F}_\mathrm{v}= 0$. Crucially, the model shows that the complex effect of vortex pinning on a rough surface can be captured by a velocity-dependent effective mutual friction, where $\gamma_0 \to \gamma_0 + \kappa\rho_s\Gamma$.  This modification can be expressed in terms of the standard mutual friction parameters $\alpha$ and $\alpha'$. The resulting effective parameters, $\hat\alpha$ and $\hat\alpha'$ become strongly dependent on the superfluid velocity $v_s$ relative to the depinning velocity $v_d$ and are shown in Fig.~\ref{fig:simulations}f for 1.3~K. The full expressions \eqref{eq:alpha-hat} are provided in the End Matter.

Finally, the numerical simulation of the vortex motion proceeds by integration of the equation of motion given by
\begin{equation}
    \label{eq:eom-pinning}
    \dot{\bv r}_k = \bv{v}_L = (1 - \hat{\alpha}^\prime)\bv{v}_s - \hat\alpha{\bv{s}'}\times\bv{v}_s,
\end{equation}
for $v_s > v_\mathrm{d}$ and $\bv{v}_L = 0$ otherwise, where $\bv{r}_k$ is the position of $k$th vortex. Here we set $\bv{v}_n = 0$ and $\bv{v}_s$ is the sum of externally applied (probe) velocity $v_p\cos(2\pi f_p t)$ ($f_p=2$~kHz for all simulations) and velocity induced by all other vortices and their images due to boundary conditions. Note that as $v_s$ approaches $v_d$ from above, $\hat\alpha' \to 1$ (and $\Gamma\to\infty$) causing the vortex velocity given by \eqref{eq:eom-pinning} to smoothly decrease to zero. For all numerical simulations reported below, the domain size was $1\times 1$~mm$^{2}$, the boundary conditions along the $x$ direction were periodic and along $y$ reflective walls at $y=0$ and $y=1$~mm. Vortices of opposite signs approaching closer than 100~nm are annihilated. See End Matter for further technical details of the simulation.

We simulate the decay of typically $2\times 10^4$ vortices with zero net circulation arranged in four types of initial conditions illustrated in Fig.~\ref{fig:simulations}(a-c): random initial distribution and vortex dipoles (visually indistinguishable), $2\times 2$, and $2\times 10$ vortex grid. The large eddies are constructed by placing equal-sign vortex points in a gaussian distribution around the eddy center with the width of 1/5 of the inter-eddy spacing. The vortex dipoles are pairs of vortices with uniformly random positions and orientations gaussian distribution of separations with 1~\textmu m mean. Without pinning (Fig.~\ref{fig:simulations}d), the decay follows $1/t$ with the prefactor depending weakly on the initial condition, with the exception of the pairs initial condition which shows rapid decay. With pinning, but without probe flow, the vortex line density saturates at a nonzero value when the superfluid velocity induced by neighboring vortices falls below the depinning velocity (Fig.~\ref{fig:simulations}d, similar behavior is seen for $v_p < v_d$ for all initial conditions). When probe velocity of amplitude $v_p$ is introduced (Fig.~\ref{fig:simulations}e, $v_p$=11 cm/s, $v_d$=10 cm/s), the decay develops non-self-similar features: a faster-than-$1/t$ decay, followed by a significant slowdown around 0.5~s after which the final $1/t$ is slowly recovered. The decay of the vortex pairs is again more rapid following approximately $t^{-2}$.

\emph{Discussion. -- } The broken self-similarity appears since the pinning and probe flow introduce special velocity scales. The fast initial decay most likely appears due to presence of small vortex pairs, whose rapid annihilation scales roughly as $t^{-2}$ as shown in Fig.~\ref{fig:simulations}(e). Once these small pairs annihilate, any further decay can proceed only after large initial clusters are mixed (compare the slowdown in the range 1 -- 10 s in \ref{fig:decay-data-all}c and Fig.~\ref{fig:simulations}e), which is further slowed down by high $\hat\alpha^\prime$ just above depinning velocity, Fig.~\ref{fig:simulations}f, which slows down advection in \eqref{eq:eom-pinning}. The slowdown is absent for the random or pairs initial conditions with the same depinning and probe velocities, although after the vortices become mixed, the decay becomes mostly independent of the initial condition. 

For low probe velocities ($v_p \approx 4$~cm/s, gray curves in Fig.~\ref{fig:decay-data-all}b), the probe signal recovers rapidly. This is consistent with the immobilization of vortices seen in Fig.~\ref{fig:simulations} for $v_p < v_d$: the superflow is dissipated by the vortices at an instantaneous power $P$ given by the Magnus force with \eqref{eq:eom-pinning}, i.e., $P = \kappa\rho_s\hat\alpha v_s^2$ and for $v_s < v_d$, $\hat\alpha = 0$ and thus $P = 0$. This, however, also modifies the applicability of \eqref{eq:L}, where the bare mutual friction parameter $\alpha$ ought to be replaced with $\hat\alpha$ averaged over one cycle of the oscillating flow, which would renormalize the measured $L$ by approximately 1/4 (see End Matter), thus changing the prefactor of the decay law, but would leave the scaling unchanged. Further, the flow enhancement near corners and stagnation points for the G and J geometries will lead to a distribution of local $v_s$ which affect $\hat\alpha$ and spatial sensitivity to $L$ (especially for \emph{J} device with strongly localized probe flow, Fig.~\ref{fig:meas_sequence}a). More work is needed for the full understanding of the interaction of pinning and dissipation, however, the numerical model captures all essential features of the experimental observations.

The change in the behavior of the decay as the probe velocity is increased can be seen as the superfluid analogy of classical (quasi-)two-dimensional flows over topography: for low turbulent energies, the vorticity becomes locked to the topography and does not decay in the inviscid limit \cite{Bretherton1976,Siegelman2023,Gallet2024}; for higher energy, the  vortices freely roam the domain, although the details depend on the initial conditions \cite{He2024}. It is tempting to connect the observed decay to the vortex gas regimes of classical 2D turbulence \cite{Mcwilliams1990,Carnevale1991,Gallet2020,Meunier2025} where the density of coherent vortices is found to decay as $t^{-0.7}$ \cite{tabeling_1991,cardoso_1994,hansen_1998}. However, these vortices change strength and size during the decay \cite{Mcwilliams1990}, and would thus correspond to clusters of quantized vortices in the superfluid. On the other hand, considering a core of a quantized vortex with finite enstrophy, e.g., a Rankine vortex with uniform vorticity in the core of radius $a_0$, the enstrophy maximum is conserved exactly due to quantization of circulation, fulfilling the assumption of \cite{Carnevale1991}, and the mean enstrophy $z  = (\kappa^2 L)/(\pi^2a_0^4)$, i.e. $z\propto t^{-1}$, in agreement with classical decay of enstrophy which follows $t^{-\nu}$ with $\nu = 0.8 $ to $1.6$ \cite{clercx_2003}. 

\emph{Conclusions. -- } We have observed the first temporal decay of quasi-2D turbulence in superfluid \4He confined in nanofluidic channels. Three different geometries show a common behavior of vortex density decay with universal fast initial transient $L(t)\approx 50 [\mathrm{mm}^{-2} \mathrm{s}^2]/t^2$ followed by a slower regime generally compatible with $L\propto t^{-1}$. This observation is consistent with the theoretically predicted $t^{-1}$ decay from two-vortex annihilation at finite temperatures \cite{baggaley_decay_2018}, while the deviations from this behavior are shown to be a consequence of vortex pinning. We demonstrate that this complex interaction with the rough walls can be successfully modeled using a velocity-dependent effective mutual friction, which reproduces all essential features of our experimental data. This strongly confined flow could be used as a model for other systems where pinning of quantized vorticity on disorder is important, such as pulsar glitches \cite{Antonelli2020}.

\begin{acknowledgments}
     The work was supported by Charles University under PRIMUS/23/SCI/017 and the Czech Science Foundation under GA\v{C}R 25-16386S. CzechNanoLab projects LM2023051 and LNSM-LNSpin funded by MEYS CR are gratefully acknowledged for the financial support of the sample fabrication at CEITEC Nano Research Infrastructure and LNSM at FZU AV\v{C}R. AFM characterization was performed in MGML laboratories supported by the program of Czech Research Infrastructures (project no. LM2023065). We also thank L. Skrbek and B. Szalai for fruitful discussions.
\end{acknowledgments}

\bibliography{apssamp}

\section{End Matter}

\begin{figure}
    \centering
    \includegraphics[width=\linewidth]{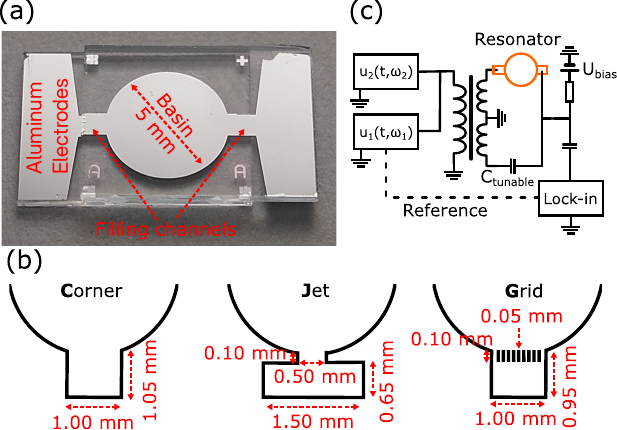}
    \caption{(a) - A photo of the Helmholtz resonator. (b) - Sketches of resonator channels with dimensions, which are then used to calculate the probe velocity $v_{probe}$. (c) - A sketch of the measurement circuit. }
    \label{fig:circuit}
\end{figure}

\paragraph{Experimental details and data smoothing. -- }A photo of the device and detailed drawing of the flow channels is shown in Fig.~\ref{fig:circuit}a,b. The response of the device is measured with the capacitance bridge shown in Fig.~\ref{fig:circuit}c. The primary winding of the transformer is excited by two generators in parallel which force the pump and probe resonances of the device wired in one of the arms of the capacitance bridge excited by the center-trapped secondary winding of the transformer. The tunable capacitor $C_\mathrm{tunable}$ is set to minimize the signal seen by the lockin, referenced to the probe tone, off resonance. For the decay measurements, the lock-in time constant is set to 10~ms with filter slope set to 12~dB/octave and the time-series is recorded into the internal lockin buffer memory at 64~Hz.

\begin{figure}
    \centering
    \includegraphics[width = \linewidth]{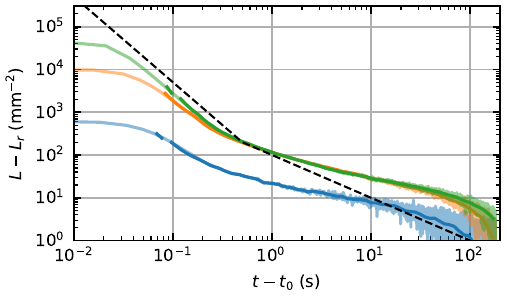}
    \caption{Decay of vortex line density in the G device. As Fig.~\ref{fig:decay-data-all}c, but for $T = 1.450$~K; $v_p = 0.16$ms$^{-1}$. The black dashed lines indicate $100/t$, $50/t^{2}$ as in Fig.~\ref{fig:decay-data-all}.}
    \label{fig:decays_G_1450}
\end{figure}

Since the area where the probe mode is sensitive to the presence of vortices is approximately 1~mm$^2$ (see Fig.~\ref{fig:meas_sequence}a), for $L<10$~mm$^{-2}$ only few individual vortices are present above the quiescent background, resulting in a significant noise level. The obtained decay curves are thus smoothed by averaging in a sliding window in logarithmic coordinates (i.e., the window is given by $t_l < t < t_r$, with $t_r/t_l=\mathrm{const.}$ and the averaged point is assigned to $t = \sqrt{t_lt_r}$) (see Fig.~\ref{fig:decays_G_1450} for typical non-smoothed noise level).

\paragraph{Derivation of Eq. 2. and simulation details. -- }
For $|\bv{v}_s| > v_d$, the vortex velocity $\bv{v}_L$ in \eqref{eq:vl} is given simply by $\bv{v}_L = \bv{v}_s + \bv{v}_i$, where $\bv{v}_i$ is the self-induced velocity. Referring to Fig.~\ref{fig:meas_sequence}e, in the local induction approximation, $|\bv{v}_i| = v_d$ and direction is parallel with $\bv{s}'\times\bv{s}''$, i.e. perpendicular to the plane of the vortex. The orientation of the plane of the vortex is such that the component of $\bv{v}_s$ perpendicular to this plane is equal in magnitude and opposite to $\bv{v}_i$. For the apex of the vortex (i.e., the point farthest from the pinning sites) this can be written as
\begin{equation}
    \label{eq:azimuthal}
    \frac{1}{v_L}(\bv{s}'\times\bv{v_L})\cdot\bv{v}_s = v_d,
\end{equation}
the vortex velocity $\bv{v}_L$ being constant along the vortex for a circular geometry in local induction approximation. Without loss of generality, we can write $\bv{v}_L = c_1 \bv{v}_s + c_2\bv{s}'\times\bv{v}_s$; substituting this ansatz to \eqref{eq:azimuthal} and rearranging yields
\begin{equation}
    \label{eq:c2}
    c_2 = -|c_1|\frac{v_d}{v_s}\left(1 - \frac{v_d^2}{v_s^2}\right)^{-1/2}.
\end{equation}
The equation for $c_1$ can be obtained from $\bv{v}_i = \bv{v}_L - \bv{v}_s$ by substituting the $\bv{v}_L$ ansatz and using \eqref{eq:c2}. The equation for $c_1$ arises from setting $|\bv{v}_i| = v_d$; assuming $c_1 > 0$ this yields
\begin{equation}
    v_d^2 = (c_1 - 1)^2v_s^2 + c_1^2\frac{v_d^2}{v_s^2}\left(1 - \frac{v_d^2}{v_s^2} \right)^{-1}v_s^2,
\end{equation}
which is a simple quadratic equation with a single solution $c_1 = 1 - v_d^2/v_s^2$ from which \eqref{eq:vl} follows.

Inverting the expression of the shifted $\gamma_0 + \kappa\rho_s\Gamma$ in terms of $B$ from \cite{Barenghi1983}, and using $B=\alpha\rhon/(2\rho)$ yields the modified $\hat\alpha$ in terms of bare $\alpha$
\begin{subequations}
\label{eq:alpha-hat}
\begin{align}
    \hat\alpha &= \frac{\Gamma \alpha^{2} + \Gamma \alpha'^{2} - 2 \Gamma \alpha' + \Gamma + \alpha}{\Gamma^{2} \alpha^{2} + \Gamma^{2} \alpha'^{2} - 2 \Gamma^{2} \alpha' + \Gamma^{2} + 2 \Gamma \alpha + 1},\\
    \hat\alpha' &= \frac{\Gamma^{2} \alpha^{2} + \Gamma^{2} \alpha'^{2} - 2 \Gamma^{2} \alpha' + \Gamma^{2} + 2 \Gamma \alpha + \alpha'}{\Gamma^{2} \alpha^{2} + \Gamma^{2} \alpha'^{2} - 2 \Gamma^{2} \alpha' + \Gamma^{2} + 2 \Gamma \alpha + 1},
\end{align}
\end{subequations}
which are shown in Fig.~\ref{fig:simulations}f for 1.3~K (for full detailed derivation see \cite{dataset}). Note that for moderate velocities at low temperatures, the effective mutual friction is almost entirely determined by the interaction with the surface.

The superfluid velocity that enters the equation of motion \eqref{eq:vl} is
\begin{equation}
    \label{eq:total-vs}
    \bv{v}_s(\bv{r}_k) = \bv{v}_p + \frac{\kappa}{2\pi}\sum_{j\neq k}\frac{\bv{s}'_j\times(\bv{r}_k - \bv{r}_j)}{|\bv{r}_k - \bv{r}_j|^2},
\end{equation}
where $\bv{r}_k$ is the position of the $k$th vortex and the sum runs over all other vortices, including the image vortices due to boundary conditions.

The boundary conditions were implemented using image vortices by either periodic repetition (along the $x$-direction) or mirror reflection (along the $y$-direction at $y = 0$ and 1~mm) with a total of 8 image domains ($\pm$ directions in $x$ and $y$ and diagonals). For time stepping, simple Euler stepping with fixed $\Delta t = 10^{-5}$~s was used (a subset of the runs was validated with $\Delta t = 10^{-6}$~s, yielding identical results). Vortex points of opposite signs approaching closer than 100~nm are annihilated and removed from the simulation. The vortex system is left to freely decay from the prescribed initial condition with no additional vortex injection. The source code of the simulation including the configuration scripts of the simulations reported here are available in the attached dataset \cite{dataset}.

\paragraph{Dissipation of fourth sound with effective mutual friction}
\label{par:alpha-bar}

\begin{figure}
    \centering
    \includegraphics{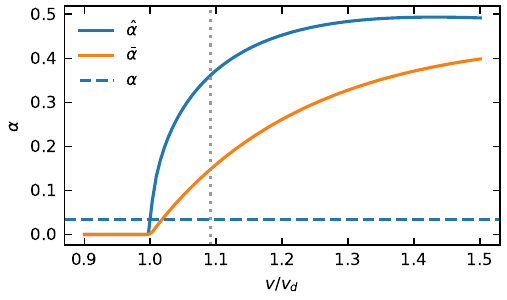}
    \caption{Bare $\alpha$, pinning-modified $\hat\alpha$ and average $\bar\alpha$ for AC flow. For $\bar\alpha$ the velocity $v$ refers to the velocity \emph{amplitude}.}
    \label{fig:mean-alpha}
\end{figure}

The quality factor $Q$ of the fourth sound resonance is given by, from definition,
\begin{equation}
    Q^{-1} = Q^{-1}_0 + \frac{1}{2\pi}\frac{\int_0^{1/f_0}\dd t\iint \dd^2S \rho_s\kappa L \hat\alpha(v_s)v_s^2}{\frac{1}{2}\iint\dd^2 S \rho_s v_s^2},
\end{equation}
where $Q_0$ is the quality factor associated with intrinsic losses, the integral in the numerator gives the total dissipated energy by the Magnus force with vortex velocity given by \eqref{eq:eom-pinning} over one cycle of the resonance and the double integral runs over the area of the resonator. Defining $\bar\alpha \equiv \int \hat\alpha(v_s)v_s^2 \dd t/\int v_s^2 \dd t$, where the integration runs over one cycle of the Helmholtz mode, results in $Q^{-1} = Q_0^{-1} + \frac{1}{2\pi}\bar\alpha\kappa L$, from which Eq.~\ref{eq:L}, with $\alpha$ replaced by $\bar\alpha$ follows. The comparison of bare $\alpha$, $\hat\alpha$, and $\bar\alpha$ is shown in Fig.~\ref{fig:mean-alpha}. For 1.3 K, $v_p = 12$~cm/s and $v_d = 11$~cm/s, the ratio $\bar\alpha / \alpha \approx 4.3$, i.e., a vortex sliding on a rough surface dissipates 4 times as much as vortex on a smooth surface.

\end{document}